# Green and Safe 6G Wireless Networks: A Hybrid Approach

Haneet Kour, *Student Member, IEEE*, Rakesh Kumar Jha, Senior *Member, IEEE, Sanjeev Jain, Member IEEE*

*Abstract*— With the wireless internet access being increasingly popular with services such as HD video streaming and so on, the demand for high data consuming applications is also rising. This increment in demand is coupled with a proportional rise in the power consumption.  It is required that the internet traffic is offloaded to technologies that serve the users and contribute in energy consumption. There is a need to decrease the carbon footprint in the atmosphere and also make the network safe and reliable. In this article we propose a hybrid system of RF (Radio Frequency) and VLC (Visible Light Communication) for indoor communication that can provide communication along with illumination with least power consumption. The hybrid network is viable as it utilizes power with respect to the user demand and maintains the required Quality of Service (QoS) and Quality of Experience (QoE) for a particular application in use. This scheme aims for Green Communication and reduction in Electromagnetic (EM) Radiation. A comparative analysis for RF communication, Hybrid (RF+ VLC) and pure VLC is made and simulations are carried out using Python, Scilab and MathWorks tool. The proposal achieves high energy efficiency of about 37%, low Specific Absorption Rate (SAR), lower incident and absorbed power density, complexity and temperature elevation in human body tissues exposed to the radiation. It also enhances the battery lifetime of the mobile device in use by increasing the lifetime approximately by 7 hours as validated from the obtained results. Thus the overall network reliability and safety factor is enhanced with the proposed approach.

## I. INTRODUCTION

With the high rise in the need and popularity of wireless communications, there is an unabated demand for ubiquitous connectivity in the network. Future generations of wireless networks are required to serve novel applications such as enhanced mobile broadband (eMBB), Ultra-Reliable Low Latency Communication (URLLC), massive Machine Type Communication (mMTC) and many more [1].

H. Kour is with Central University (CU), J&K, India., R. K. Jha is with IIIDM Jabalpur, India  and Sanjeev Jain is with CU J&K (E-mail:, hani.kpds@gmail.com, jharakesh.45@gmail.com, dr_sanjeevjain@yahoo.com )

The wireless communication industry is faced with rising challenges because of increasing complexity and rise in the count of devices accessing the enhanced applications. Various architectural designs and efficient technologies have gained popularity as potential solutions to utilize the available spectrum efficiently and accommodate more users. The future cellular networks are expected to incorporate energy efficient techniques such as device-to-device communication (D2D), spectrum sharing (SS), expanding the range of the spectrum (mmWave i.e. Millimeter wave communication), small cell access points. Other opportunities include Massive MIMO (Multiple Input Multiple Output), beamforming, NOMA (Non-Orthogonal Multiple Access), Mobile Edge computing (MEC). Visible Light Communication (VLC) is a potent solution to be integrated within 5G networks because of high bandwidth density ($b/s/m^2$) as available in optical signals. Integration of VLC and 5G can help in overcoming the limitations of each of the technologies to maximize the total aggregate throughput achieved [2]. The various advantages associated with VLC include availability of unregulated spectrum, high energy efficiency and security of the network. Since visible light is not regulated, VLC can prove to be a cost effective solution for the wireless communication industry. Also, the indoor environments are mostly illuminated which makes visible light communication to be most advantageous in indoor communication environments. Additionally, VLC can also aid in reducing the carbon dioxide footprint produced from the ICT (Information and Communication technology) industry as there is no RF spectrum involved and hence no Electromagnetic pollution due to light communication.

VLC can aid in minimizing the EM (Electromagnetic) radiation effect in the atmosphere resulting because of RF communication in the network. Integration of VLC with RF communication can provide a potential solution in achieving high data rate and reliable communication for indoor communication scenarios. Since most of the times a device user is operating in indoor environments such as homes, offices and so on, VLC can be deployed in such scenarios with the aim of decreasing the congestion in the network and reducing the load on RF cells. Since VLC is LED based illumination, it cannot cause EM radiation impact to the users. The light intensity in a VLC system can be regulated by various dimming techniques such as OOK (On Off Keying) dimming or MPM (Mirror Pulse Modulation) dimming [3]. If there is proper support maintained for visibility and flicker mitigation during the active and idle periods at the infrastructure of VLC, it can result in high service requirements to be achieved in a communication system. A comparison between RF and VLC communication is tabulated in Table I.

TABLE I
COMPARISON TABLE

| S.No. | Parameters | RF | VLC |
|---|---|---|---|
| 1. | Transmitter | Base Station, SCA | LED/ Laser Diode (LD) |
| 2. | Receiver | Receiver Antenna | Photodiode / Camera |
| 3. | Distance | 500 m- Km | 20 m |
| 4. | Electromagnetic Interference | Yes | No |
| 5. | Data Rate | 100-200 Mbps (4G LTE) | 10 Gb/s using LED 100 Gb/s using LD |
| 6. | Frequency | 30-300 Hz or in MHz range | 430 THz -790 THz |
| 7. | Spectrum | Radio Frequency | Visible light |
| 8. | Security | Limited | High |
| 9. | Coverage | High | Limited |
| 10. | System Complexity | High | Low |
| 11. | Transmission Power | Dozens of watts | Few Watts |
| 12. | LOS/NLOS | Both | LOS |

*A. Related Work*

Optical Wireless Communication (OWC) makes use of the visible light, infrared and ultraviolet for propagation as the wireless media. There are various systems that use OWC technologies such as Li-Fi (Light Fidelity), VLC, FSO (Free Space Optical) communication, OCC (Optical Camera Communication) being the most popular and promising OWC technologies [4]. VLC was first proposed in 1999 and several standards for the same have been released such as VLCC-STD-001, JEITA CP-1221, JEITA CP-1222 and IEEE 802.15.7 [5]. The standardization activities for VLC have been tabulated in Table II. Visible Light Communication (VLC) can form a major part of the indoor wireless communication as the indoor communications form the majority of the traffic in the upcoming 5G and future communication networks [6-7]. It has been reported from the available data that about 70% of the internet traffic occurs indoors and only 30% occurs in outdoor scenarios [8]. Integrating VLC and Radio frequency can provide a potential solution for the "Wi-Fi Spectrum Crunch" problem with the use of low cost light emitting technology such as LED's. RF and VLC when integrated can provide increased coverage to ensure stable data rates. LED's can be used for energy efficient wireless access to provide illumination as well as wireless communication [8].

Various network architectures have been presented in literature that offer high speed VLC for a local area network based on different network topologies [9]. The design of the VLC system is done as to conserve maximum energy with optimum LED illumination and communicate with the available modulation schemes. VLC finds applications in Indoor Communication, Vehicular communication, Underwater communication along with the respective challenges in each. As RF and VLC do not interfere with each other, they can co-exist and be operational in the same indoor areas such as offices, indoor campus areas of educational institutions.

TABLE II
STANDARDIZATION ACTIVITIES FOR VLC.

| S.No. | VLC Standard | Year |
|---|---|---|
| 1. | JEITA CP-1221 | 2007 |
| 2. | JEITA CP-1222 | 2008 |
| 3. | VLCC-STD-001 | 2008 |
| 4. | IEEE 802.15.7 | 2011 |
| 5. | ITU g.hn | 2011 |
| 6. | IEEE 802.15.7 | 2018 |

The main challenges and open issues in VLC include Line of Sight (LOS) communication, flickering, noise, interference, dimming and mobility [10]. The sharp decrease in achieved data rate as the link distance increases in VLC is a challenge. VLC requires specialized high speed photo diode receivers as well as highly accurate 3D positioning and orientation. Another challenge is the Commercialization of VLC as it requires the lighting Original Equipment manufacturers (OEMs) to make modifications in the lamps/fixtures designing. The existing system is high in complexity which can be reduced by incorporating Hybrid-VLC communication for indoor scenario.

VLC deployed in small cells such as for indoor communication can aid in significantly increasing the network capacity, prolonging/ enhancing the battery life of the mobile terminal whilst expanding coverage [5]. The motivation for this work arises from one of its most appealing advantages i.e. no harmful radiation impact on health making it biological safe for operation. Since radio waves can be a potential cause for causing health hazards including cancer in human beings [11-15], there is a need to make the network safe in operation as well as reliable for humans operating in close vicinity with mobile and wireless devices. This work presents the improvement in biological safety of the network incorporating VLC studied in terms of Electromagnetic radiation metrics. The "Internet of Light" is the future of internet owing to the requirements and expectation from future wireless communication networks.

*B. Contributions and Organization:*

In this article, a hybrid wireless network that integrates RF and VLC communication is proposed as a step towards Green communication to make future networks safe and reliable. A mathematical model for the proposed Hybrid RF-VLC method is presented and the simulations are carried out using Python,

Scilab and MathWorks tool to study the Electromagnetic radiation exposure. The performance of the proposed methodology is studied in terms of EM radiation metrics such as Power density, Specific Absorption Rate, Temperature elevation and energy efficiency.

Major contributions of the article include:
- This paper presents a hybrid technique for indoor wireless access as it occupies maximum part of the communication and can be made safe and reliable in operation.
- The proposal is given to have high QoS and QoE for indoor communication and with increased security of data.
- There is comparison of EM radiation impact that is produced by current communication scenario and its improvement with the proposed hybrid approach.
- It is inferred from the proposal that it is highly energy efficient, very safe and reliable for all indoor communication scenario as validated form achieved results.
- The proposed method also aids in augmenting the battery lifetime of a device in use and hence prolongs the number of hours the battery of the device lasts.

The rest of the paper is organized as follows. General Architecture for OWC is given in Section II. Section III presents Mobility management in OWC. System model is presented in Section IV with mathematical modelling for VLC and RF system. The simulations and discussions are given in Section V of the paper. Finally, the paper concludes in Section VI.

## II. GENERAL ARCHITECTURE FOR OWC

An optical wireless communication system consists of a transmitter module and a receiver module that are separated from each other physically but connected via the visible light communication (VLC) channel [3]. The transmitter contains the information in the form of binary data. It is responsible for illumination and transmitting the data by modulating the LED produced light. The encoder modulates the message signal with different modulation schemes such as OOK (On-Off Keying), VPPM (Variable Pulse Position Modulation), CSK (Color Shift Keying), MPM (Mirror Pulse Modulation). OOK indicated the presence of the digital data as ON and the absence of it as OFF. VPPM is a modulation scheme that allows controlling the width of the pulse for light dimming or to eliminate intra-frame flicker in the light. It helps in maintaining a continuous stream of data on a varied range by adjusting pulse width. CSK is a modulation scheme that is used when the VLC system involves multiple light sources. It helps in maintaining the average emitted optical power and the total optical power to a constant value during entire communication. The encoder is responsible for commanding the LED switching with respect to the digital information and the data required. MPM utilizes dimming by compensation symbol insertion, amplitude dimming, and out-of-band dimming. Additionally, mirror pulse position modulation (MPPM) utilizes pulse width dimming. Since MPM and OOK modulation are always transmitted with a fixed brightness within a symbol (such as an MPM symbol or symmetric Manchester symbol), the data frame may need to have compensation time added in order to change the observed source's average intensity. MPM dimming is used to increase average brightness (AB) by adding compensation symbols (CS).

The OWC receiver module consists of a photo sensitive element such as a photo diode or an image sensor. The receiver extracts the message signal or the binary information signal from the received modulated light. There is line of sight (LOS) communication required to be maintained between the transmitter and the receiver, as light cannot traverse the obstacles. The decoder is based on a microcontroller or FPGA (Field Programmable Gate Array). The performance of the OWC receiver affects the system performance. The communication range and interference gets affected with the performance of the VLC receiver. The performance of the receiver is required to be maintained because there is interference caused by other light sources which acts as noise. To improve the performance the receiver consists of an optical filter. The optical filter rejects the noise components which are noise such as IR components. The interference can be reduced if the field of view (FOV) of the optical system is narrowed. For long range applications narrow FOV is a solution as it provides high SNR due to concentration of light on the photodetector.

A biconvex lens is used that concentrates the light on the photodiode or photodetector and improves the performance. The generated photocurrent is converted to voltage by the trans impedance circuit. The voltage obtained is amplified and is filtered to remove the high and low frequency noise as well as the DC component. There is a data decoding unit that reconstructs the original digital message. The OWC architecture is also depicted graphically in parallel with the physical architecture in terms of the layers and the sublayers involved in Fig.1. It consists of the physical layer, MAC (Medium Access Control) layer, upper layers consisting of the network layer, application layer. These layers are defined in the IEEE Std 802.15.7 and every layer is accountable for one portion and offers the services to the succeeding layers of the architecture [3]. The physical layer consists of the light transceiver; MAC sublayer is responsible for providing a path for various types of transfers. There is also a device management entity (DME) support in the architecture. It communicates with the physical layer management entity (PLME) and MAC sublayer management entity (MLME) to interface them with the dimmer. The DME also controls the physical switch to select the optical media i.e. optical sources and photodetectors.

The physical switch is interfaced with the optical service access point and it connects the optical media and photodetectors. The physical layer III supports various optical sources for OWC cell. Data streams of order n×m, where n represents the number of cells and m by the count of data streams from the physical layer which connect the physical switch and optical-SAP (Service Access Point). In PHY III for m=3, the cells from 1 to n are represented by Cells #1 to Cell #n The devices within the cells are capable of communicating via hybrid, pure RF, or pure VLC modes. We depict cell# n-1 in which pure VLC communication is performed. In cell# n there is pure RF communication performed. There is a region where both the cells interfere. This is the hybrid mode of communication area. In this area of the cell, the received light is not of optimum intensity as it is at the edge of the cell so the received RF signal performs better. Hence RF communication will be performed at some points and VLC at other instants. This happens when the user is a mobile user due to which the intensity of the received signal changes which change in location while moving. This concept is further explained in the following Section III i.e. Mobility Management in OWC.

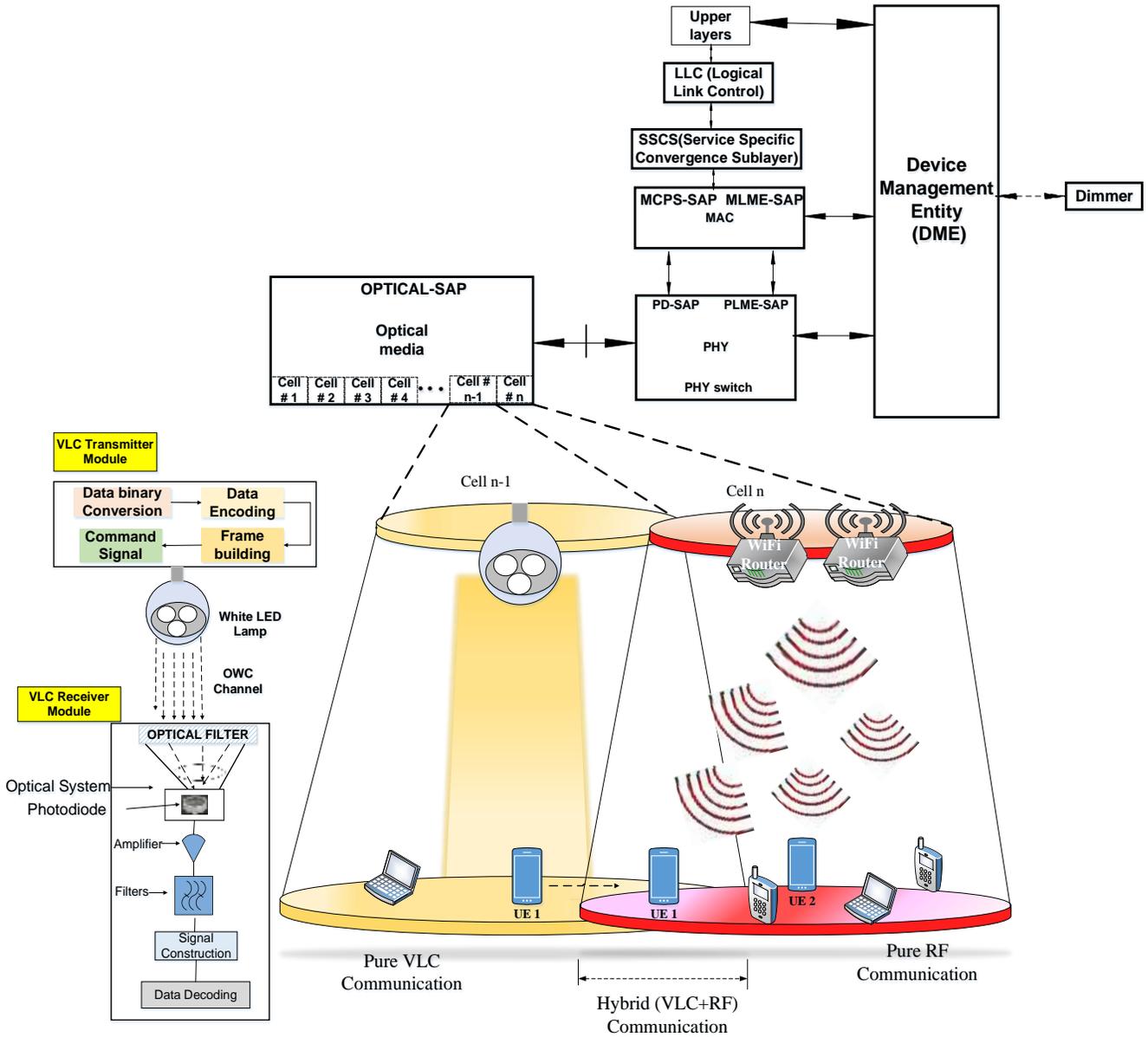

Fig. 1 General Architecture for Optical Wireless Communication.

### III. MOBILITY MANAGEMENT IN OWC

In OWC when there is a presence of a new user and there are no time slots available to be assigned then the coordinator uses multiple bands to extend the services to the new user. Fig. 2 shows the multiple channel resource assignment for OWC mobility. When a new user tries to access but no time slot is available the coordinator can assign another band to the new user. The slotted Random Access Algorithm is employed before transmitting any information or MAC command frames within the CAP (Contention Access Period), till the time the frame can be transferred soon after a data request command is acknowledged or periodic beacons are used. [3]. Multiple bands can also be used by the coordinator in which case the coordinator sends the "Src_multi_info" signal to the device in the MAC command. The device also responds with the "Des_multi_info" to inform about the multiple bands. Mobility support in OWC cell can be of two types depending on the need of physical movement or link switching in the cell.

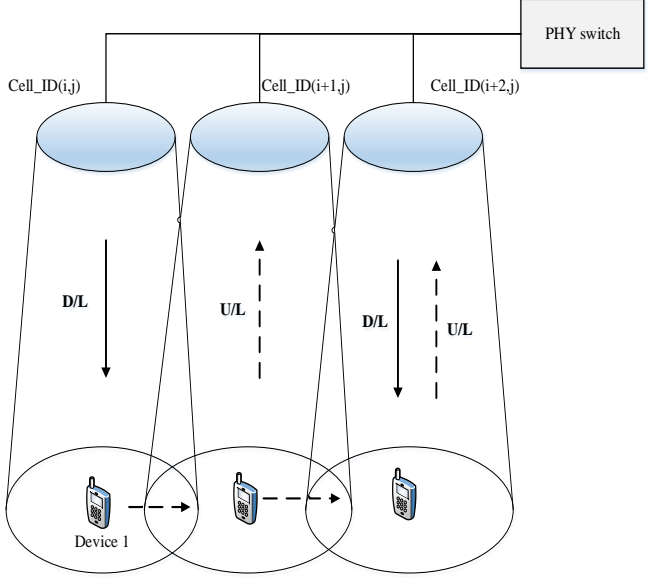

Fig. 2 Multiple channel resource assignment for OWC mobility.

The two types of mobility are: physical mobility and logical mobility. Physical mobility is when there is a change in physical position of the mobile in the coverage area of an OWC cell. Logical mobility is when the communication link changes from one to another because of interference issues or deliberate switching of the channel. Mobility in an OWC channel can be supported by a PHY switch which is further controlled by the Device Management Entity (DME). The optical elements in an OWC cell are demoted by Cell_ID (i,j) where "i" denotes the cell and "j" depicts the index of the ith elememt in the cell. When a device in a cell moves from one cell to another i.e. the mobility of the device changes, its movement is detected by the coordinator with the help of the uplink signal i.e. the acknowledgement frame.

When a device changes its position from one cell to another in the event of mobility the coordinator may or may not receive its uplink information. In such a case the coordinator searches for nearby devices in neighboring cells such as Cell_ID(i+1,j) and Cell_ID(i-1,j). Fig. 2 depicts multiple channel resource assignment for OWC mobility. The coordinator can expand its coverage by expanding the cell size to accommodate mobile users and can also decide for selecting new cell for the device on receipt of its uplink information.

## IV. SYSTEM MODEL

The system model presents the methodology of the proposed architecture for carrying out RF, VLC and Hybrid RF-VLC communication in the network. This section consists of the mathematical modeling with entire mathematical derivation of the proposed scheme. The handover mechanism for switching between RF and VLC is also studied. Also, a real case study analysis has been done in the institute campus with all the three mode of communication (as in Fig. 3 System Model). The authors have studied the biological impact i.e. the safety and reliability of the network in terms of EM radiation metrics i.e. Specific Absorption Rate, Power density, Temperature Elevation, Energy efficiency and also Battery lifetime of the mobile device. In the system model we consider an OWC network containing a macrocell (with base station serving the users) in which a university campus is further considered for studying the aspects of communication i.e. Wi-Fi communication and VLC Communication in the network.

It is assumed that the campus contains "R" Wi-Fi Access Points (AP's) and "V" VLC AP's. The VLC AP's have Light Emitting Diodes (LED's) that are responsible for illumination as well as communication. There is free space wireless communication performed with the light signals for data transmission. The Wi-Fi AP's in the network communicate over the RF signals for transmission and reception of data. The UE's that are communicating in the model are equipped with transceivers that can enable communication for both VLC and Wi-Fi. Fig.3 presents the diagrammatic representation of the system model. In the macrocell coverage area there is only radio frequency communication between the macrocell base station and small cell access points or mobile users. The university campus depicts Hybrid communication with Wi-Fi and VLC communication both. The third part depicts internal rooms in a department building where there is pure VLC communication. Our main aim is to reduce the total power consumed on a device level and network level whilst maintaining a required QoS (Quality of Service) and QoE (Quality of Experience) of the operating users. We are mostly considering indoor communication in this scenario to study the effect of VLC Communication, Wi-Fi communication and Hybrid (VLC and Wi-Fi) communication in the network.

### A. Mathematical Modelling

#### 1) VLC Radiation pattern and path loss

In a visible light communication system, there is On-Off modulation keying used. The VLC system consists of VLC Access points comprising a LED Lamp with several LED's. We consider Line of sight (LOS) communication in a VLC cell called as attocell. The active users operating under one LED lamp use the same bandwidth and the co-channel interference is considered as noise. A number of active users operate under one access point. The modulation scheme used is TDMA (Time Division Multiple Access) and RR (Round Robin) scheduling. There is a required level of illumination that has to be maintained in VLC communication. This optical power can be produced by DC current, AC current or by both.

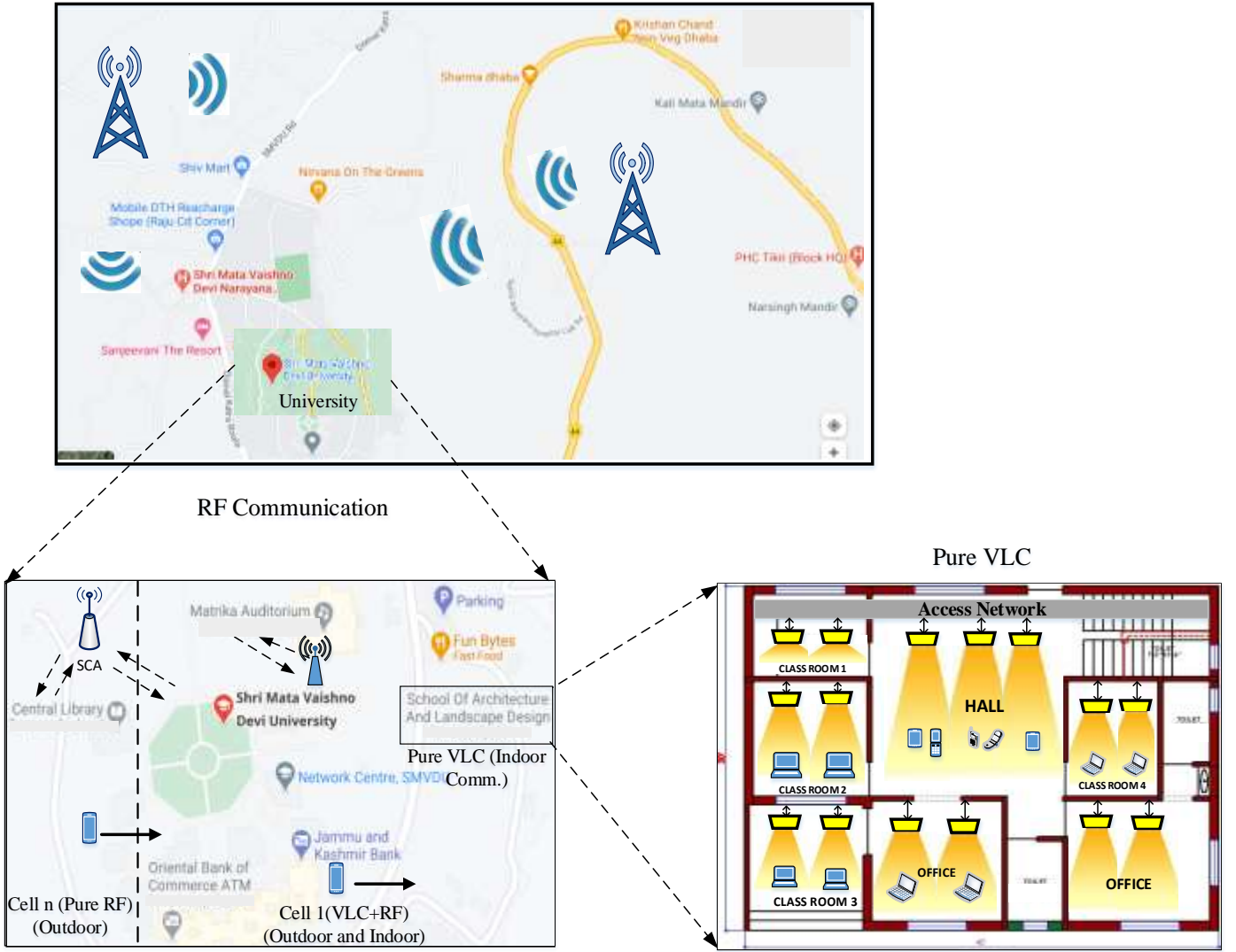

Fig.3. System Model. *(Case Study on university campus map) (Credit: Google Map).*

AC doesn't lead to flickering of light as the current changes at such a fast pace that it can't be perceived by human eye. In the DC component, the switching can lead to reduction in efficiency and hence more power consumption. If we assume "v" to be a VLC AP, then $P_{vlc}^{on}$ is the value of the consumed electrical power by the AP when DC current is used to generate the optical power. In this case $P_{vlc}^{on}$ is the total power that is consumed by the AP. If data rate is also supported by it then the power generated by AC current is used additionally represented by $P_{data}^{v}$. Hence both $P_{vlc}^{on}$ and $P_{data}^{v}$ contribute additionally when a VLC AP performs illumination and communication. The VLC's alignment angles are because of radiation patterns. In VLC Lambertian model the half power angle is half of the maximum power. The irradiance angle is the angle between the axis perpendicular to the LED surface and the direction to the receiver. The irradiance angle is represented by $\emptyset$. The incidence angle is the angle between the axis perpendicular to the photodetector surface and the direction to the transmitter. It is represented by $\theta$. The optical channel gain of the LOS channel is represented by H(0) [8].

$$H(0) = \begin{cases} \frac{(n+1)A_{pd}}{2\pi D^r} T_{of}(\theta)G(\theta)Cos^n(\emptyset)Cos(\theta) & \theta \leq \theta_F \\ 0 & \theta > \theta_F \end{cases} \quad (1)\,[21]$$

where $n$ is the Lambertian index or the order of the model. It is a function of the half-intensity radiation angle $\emptyset_{1/2}$.

$$n = -\frac{1}{\log_2 Cos(\emptyset_{1/2})} \quad (2)$$

where $r$ denotes the path loss exponent, $\emptyset$ denotes the irradiance angle and $\theta$ the incidence angle. $A_{pd}$ represents the physical area of the photodiode. $\theta_F$ depicts the half angle of the receivers Field of View (FoV). The gain of the optical filter is presented by $T_{of}$. Here $D$ stands for the standoff distance between the VLC access point and the optical receiver.

The gain of the concentrator is given by $G(\theta)$ as follows:

$$G(\theta) = \begin{cases} \frac{n_{rf}^2}{Sin^2\theta_F} & 0 \leq \theta \leq \theta_F \\ 0 & \theta > \theta_F \end{cases} \quad (3)$$

where $n_{rf}$ depicts the refractive index for a given user. The SINR (Signal to Interference plus Noise Ratio) for a user $j$ connected to a VLC access point $k$ is given as follows:

$$SINR_{j,k} = \frac{(\gamma P_{op} h_{j,k})^2}{N_0 B_j + \sum_{x \neq k}(\gamma h_{jx} P_{op})^2} \quad (4)$$

where $\gamma$ represents the optical to electric conversion coefficient. $P_{op}$ is the optical power transmitted by the VLC access point. $N_0$ is the noise power spectral density. $h_{j,k}$ is the channel gain between the user $j$ and access point $k$. $h_{j,k}$ is the channel gain between the user $j$ and an interfering VLC access point $x$. In a VLC system optical OFDM (Orthogonal Frequency Division Multiplexing) is employed. The modulation technique in a VLC system is Intensity Modulation (IM) and Direct Detection (DD) due to which only the real values of the signals are received by the receiver. This exploits only half of the bandwidth and the achieved data rate is given as:

$$R_j = \frac{B}{2} log_2(1 + SINR_j) \text{ bps} \quad (5)$$

The SINR is computed between a user $j$ and access point $k$. Let $U$ be the total count of users designated to the $k^{th}$ access point, then the data rate is given as follows

$$R_j = \frac{B}{2U_k} log_2(1 + SINR_j) \text{ bps} \quad (6)$$

where $U_k$ are the total users designated to the $k^{th}$ VLC access point.

*2) Radio frequency system*

In the RF system there is power required to turn ON the Access point and additional power is required to support data rate such as in the case of VLC. Here $P_{wifi}^{on}$ represents the power required by Wi-Fi AP to be ON. The extra amount of power required to support data rate is represented as $P_{data}^r$. For our study of the RF system we assume a wideband RF channel to accommodate users with high bandwidth demanding applications. We assume that the system employs OFDM modulation scheme with the spectrum allocated having $B_{RF}$ bandwidth and total transmission power as $P_{RF}$. The channel gain for $m^{th}$ RF access point to $i^{th}$ in the $q^{th}$ sub-channel is given as follows

$$h_{im}^q = \sqrt{10^{\frac{-L(d_{im})}{10}}} h_{w,im}^q \quad (7)$$

The normalized channel transfer function from $m^{th}$ RF access point to the user $i$ in the $q^{th}$ sub-channel is depicted as $h_{w,im}^q$. Here $L(d_{im})$ represents the large scale fading loss in the channel in dB. The Signal to Interference plus Noise Ratio received at the $i^{th}$ user when it is separated at a distance of $d_{im}$ is given as

$$SINR = \frac{|h_i^q|^2 \Delta P_i}{N_0 \Delta B_s + I_i} \quad (8)$$

The data rate received at the $i^{th}$ user is given as

$$R_i = \sum_{q \epsilon Q_i} \Delta B_s log_2 (1 + \frac{|h_i^q|^2 \Delta P_i}{N_0 \Delta B_s}) \quad (9)$$

Where $Q_i$ is the set of sub channels for a particular user $i$. The sub channel bandwidth is denoted as $\Delta B_s$. The power allocated to each of the sub channel is denoted as $\Delta P_i$. The RF scenario consists of $U_r$ number of users assigned to $m^{th}$ RF access point, i.e. $U_r = \sum_i U_{ri}$.

The achievable rate in a scenario where the users are distributed randomly and spectrum is allocated based on the demand and user traffic is given as follows:

$$R_i = \frac{B_{RF}}{U_r} log_2 \left(1 + \frac{|h_i|^2 P_{RF}}{N_0 B_{RF}}\right) \text{ bps} \quad (10)$$

*3) Case Study*

**Case 1:** In a wireless communication system lasting for 1 hour, there is VLC communication for 40 minutes and RF communication for 20 minutes. Let $p$ be a time factor for which there is VLC communication and $q$ the time factor for RF communication. In this case, $p > q$. This is an indoor communication scenario and the user is communicating in a closed space/room for most part of the communication. VLC communication lasts for 67% of the entire duration and RF for 33% of the duration. There is VLC communication in the system when the data rate to be achieved is high and LOS communication is possible. Otherwise, there is RF communication in the system.

The data rate achieved for the duration a device is communicating in VLC is given as

$$R_{vlc} = \left[\frac{B}{2} log_2(1 + SINR_j)\right] \times p \quad (11)$$

$$R_{vlc} = \left[\frac{B}{2} log_2(1 + SINR_j)\right] \times 67\% \quad (12)$$

When the device communicates using RF communication, achieved data rate is given as

$$R_{RF} = \left[\Delta B_s log_2 (1 + \frac{|h_i^q|^2 \Delta P_i}{N_0 \Delta B_s})\right] \times q \quad (13)$$

$$R_{RF} = \left[\Delta B_s log_2 (1 + \frac{|h_i^q|^2 \Delta P_i}{N_0 \Delta B_s})\right] \times 33\% \quad (14)$$

Since $p > q$ in this case, the data rate achieved $R_{vlc} > R_{RF}$.

**Case 2:** In a wireless communication system lasting for an hour, there is RF communication for 40 minutes and VLC communication for 20 minutes. In this case $q > p$. This is a scenario where most of the time the user is communicating in an outdoor scenario or there are frequent handovers. Since a LOS connection is not easy to be maintained in such a scenario so there is RF communication for most part of the communication. Here the data rate achieved with RF is higher than that achieved with VLC because of the former lasting for a higher duration.

The achieved data rate for VLC communication lasting for 20 minutes i.e. 33% of the entire duration is as follows:

$$R_{vlc} = \left[\frac{B}{2} log_2(1 + SINR_j)\right] \times 33\% \quad (15)$$

The data rate achieved with RF communication lasting for 40 minutes i.e. 67% of the entire duration is as follows:

$$R_{RF} = \left[\Delta B_s log_2 (1 + \frac{|h_i^q|^2 \Delta P_i}{N_0 \Delta B_s})\right] \times 67\% \quad (16)$$

Since $q > p$ in this case, therefore $R_{RF} > R_{vlc}$.

**Case 3:** In this case, the VLC and RF communication last for equal duration of time. For a communication scenario lasting for 1 hour i.e. 60 minutes, the VLC communication and RF communication both last for 30 minutes each. In this case $p = q$. This is a scenario where there is equal amounts of indoor and outdoor communication happening in the system. There is LOS as well as NLOS communication in the system. The data rate achieved with VLC and RF communication in this case depends on the connection that can be established with the user. If a LOS connection is possible then there is VLC communication otherwise there is RF communication in the system.

The data rate achieved for VLC communication lasting for 30 minutes i.e. 50% of the duration is as follows:

$$R_{vlc} = \left[\frac{B}{2} \log_2(1 + SINR_j)\right] \times 50\% \quad (17)$$

The data rate achieved with RF communication lasting for the other half i.e. 30 minutes is given as:

$$R_{RF} = \left[\Delta B_s \log_2\left(1 + \frac{|h_i^q|^2 \Delta P_i}{N_0 \Delta B_s}\right)\right] \times 50\% \quad (18)$$

### B. Power Consumption and Radiation Reduction

The amount of power consumed, power saved and the EM radiation impact with VLC, Wi-Fi and Hybrid communication is studies. We have calculated Energy Efficiency and battery lifetime for evaluating power conservation as well as Specific Absorption Rate (SAR), Temperature elevation as radiation metrics for EM radiation effect. These parameters were also calculated in our recently published work [16-17] with an aim to reduce harmful EM radiation impact in 5G and beyond networks. The energy efficiency for a device communicating with a particular application such as video data can be computed as follows for VLC Communication:

$$EE_{vlc} = \frac{Data^{appl.}}{P_{vlc}^{on} + P_{data}^v} \quad (19)$$

The energy efficiency for a device communicating with Wi-Fi communication is computed as follows:

$$EE_{wifi} = \frac{Data^{appl.}}{P_{wifi}^{on} + P_{data}^r} \quad (20)$$

Whenever a mobile device is operating i.e. either transmitting or receiving data, there is production of heat in the device circuitry due to static and dynamic power consumption. This heat is responsible for increasing the SAR of EM radiation in the exposed body tissue. The SAR computation can be done with the following expression

$$SAR = \frac{P_{exposed}}{M} \quad (21)$$

where $P_{exposed}$ is the incident power at the exposed body tissue and $M$ is the mass of the tissue in Kg. The SAR is caused due to temperature elevation produced in a tissue. This temperature elevation can be studied by bioheat equation that incorporates various parameters such as blood flow to find out the rise in temperature in degree Celsius. The detailed proof is given in Lemma 2 in [17].

The computation of Power Density (PD) radiated by wireless device at '$d$' distance from the radiating antenna can be written as

$$PD = \frac{G_{tr} P_T}{4\pi d^2} \quad (22) \quad [17]$$

where $G_{tr}$ represents the gain of the transmitting antenna, $P_T$ is the total input power to the transmit antenna and '$d$' is the distance between the radiating antenna source and the receiver.

The battery lifetime enhancement for a device can be studied with the improvement in Energy Efficiency. We assumed a device with 5000 mAh battery power and $E_{battery} = 5.45\ Wh$ [18]. i.e. $I_{battery} = 5000$ mAh. To compute the energy required by UE1 to transfer data to UE2 can be calculated by

$$E_{UE12} = \frac{E_{battery}^t}{3600}\ [Wh] \quad (23)$$

where '$t$' is the file transfer latency in seconds.

Battery Lifetime is computed by the following expression:

$$Batt_{life} = \frac{I_{battery}}{I_{UE12}} \times 0.70\ [hours] \quad (24)$$

where $I_{UE12}$ is the value of the load current that is drawn by UE1 to transfer a file to UE2. Here 0.70 is the factor that includes some external events affecting the battery life of a device.

## V. SIMULATIONS AND DISCUSSIONS

This section contains the analytical results obtained after comparing the performance of a pure VLC system with RF communication (here Wi-Fi) and a hybrid system (VLC to Wi-Fi and vice versa). The simulation parameters for VLC and Wi-Fi are tabulated in Table III and Table IV respectively. The comparison between the three types of wireless communications i.e. VLC, Wi-Fi and Hybrid network (VLC + Wi-Fi) is obtained on the basis of SINR, energy efficiency, battery lifetime, temperature elevation and Specific Absorption Rate (SAR) in human body tissue. All the simulations are carried out for an indoor communication system. We consider the third floor of Electronics and Communication Engineering Department building at our university where the 5G and IoT Lab is situated.

TABLE III
VLC PARAMETERS

| | |
|---|---|
| $P_{vlc}^{on}$ of VLC Access Point | 15 Watt |
| Semi-angle at half-power | $30^0$ |
| VLC channel bandwidth | 100 MHz |
| Constant Gaussian Noise | $4.7 \times 10^{-14}\ A^2$ |
| Detector area of photodiode | $1.0\ cm^2$ |
| O/E conversion efficiency | 0.54 A/W |
| Gain of Optical Filter | 1.0 |
| Refractive index of lens | 1.5 |
| FOV of receiver | 90 deg. |
| LED luminosity efficacy | 150 lm/W |
| DC efficiency factor | 0.1 |

TABLE IV
WiFi PARAMETERS

| | |
|---|---|
| $P_{Wifi}^{on}$ of WiFi Access Point (AP) | 10 Watt |
| WiFi Channel bandwidth | 2 MHz |
| Carrier Frequency | 2.4 GHz |
| Max. power consumption of AP | 14 Watt |
| Noise Level | -90 dBm |
| Efficiency Factor | 0.1 |

The third storey has 6 rooms located externally, 2 rooms present internally. Location of the stairways is in the corners of the building. It is also assumed that the User Equipments (UE) are uniformly present randomly in different rooms of the building. The third floor consists of total 6 Wi-Fi Access points with 4 Access points (AP) at the corners and the remaining 2 in the corridors, with all of them being equally spaced. For VLC Communication we consider the 5G and IoT lab where a light source locating at the ceiling of the lab is equipped with 4 VLC Access points.

In this work we first compute the SINR for all the considered schemes i.e. VLC, Wi-Fi and Hybrid. The application demanded is assumed to be the same for all three i.e. Video data so the throughput requirement is equal for all three. The graph obtained for SINR v/s distance is given in Fig.4.

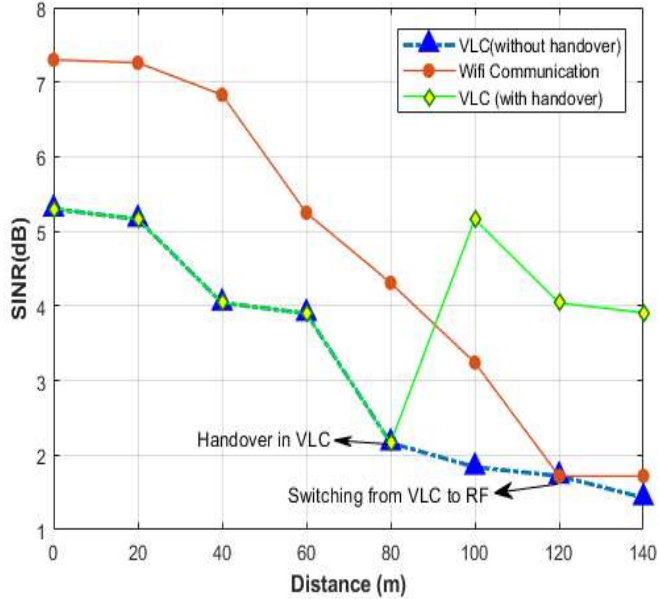

Fig.4 SINR v/s Distance for VLC (with and without Handover) and WiFi Communication.

The graph depicts the SINR obtained w.r.t distance for the mentioned schemes. As the distance increases, the SINR obtained decreases for all. It is seen from the graph that in case of VLC the SINR falls below 2dB after 80m distance and hence the required application demand can't be served. We propose handover in VLC after 80m distance. When there is handover to another VLC AP that is in close proximity to the UE, the SINR improves significantly. After 120m the SINR drops significantly in VLC communication such that no application can be served any further. At this point we propose switching from VLC to Wi-Fi as the SINR obtained in Wi-Fi is higher than that obtained in VLC communication. Next we find the power consumed during VLC and Wi-Fi communication for the same throughput demand.

For the same target data rate the efficiency of both the communication schemes is compared. The graph obtained is shown in Fig.5. It is found that as the distance increases, energy efficiency decreases for both VLC as well as Wi-Fi communication. As it has been seen that with distance SINR decreases, this decrease is SINR is responsible for lower data rates achieved with increase in distance. The power consumption for both the schemes depends on the power required for the AP to be ON and the power consumed to serve a particular application. It has been found that the total power consumption( in Watt) is higher in case of VLC AP than Wi-Fi AP but the data rate achieved in case of VLC is much higher than that obtained in Wi-Fi communication. In VLC communication the data rate obtained is of the order of 100 Mbps but in Wi-Fi communication maximum data rate obtained in our scenario is 40 Mbps.

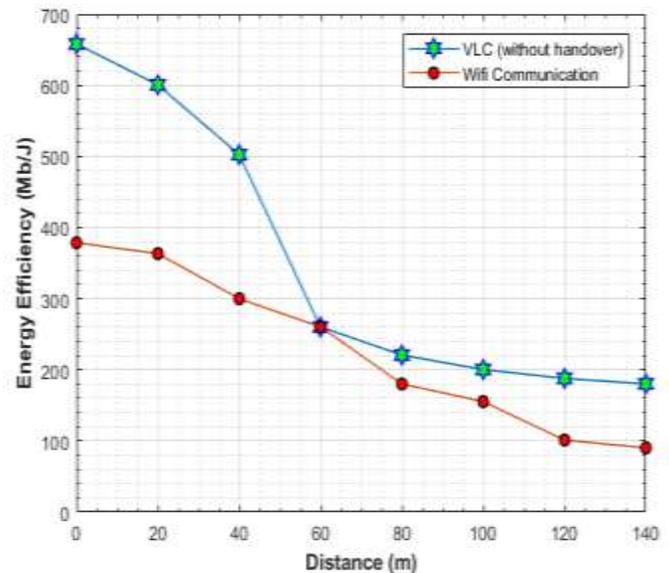

Fig.5 Energy Efficiency v/s Distance for VLC and WiFi Communication.

Since the data rate obtained in VLC communication is much higher than that obtained in Wi-Fi communication, the energy efficiency is proportionally higher. It can be seen from the graph that at a distance of 60m the energy efficiency obtained for both VLC and Wi-Fi is almost same. This is because the data rate achieved at this distance and the power consumed in each of the schemes to achieve the respective data rates is almost equal. After 60m the difference is energy efficiency for both VLC and Wi-Fi is less as compared to the difference obtained before 60m. This is due to the fact that both the schemes of communication have their respective ranges in which the performance of both is highest. Wi-Fi performs best within a range of 45-50m and VLC gives maximum performance till a range of 70-80m. To analyze the impact of EM radiation in the above discussed schemes, we use two radiation metrics i.e. temperature elevation in degree Celsius and SAR. As discussed earlier that a body tissue exposed to a amount of incident power due to a device operating causes rise in temperature of the tissue. The temperature elevation produced depends on the heating effect that arises due to light communication and radio frequency communication. This is studied when there is continuous transmission and reception of data going on in the device used such as mobile phone, with the application assumed in process being video application.

The temperature elevation has been recorded by using Pennes bioheat equation. The 3D thermal radiation pattern obtained is given in Fig.6. It depicts the rise in temperature in the human tissue with respect to skin depth and the time for which the exposure has been recorded. The 3D radiation pattern depicts a temperature rise in degree Celsius when studied on a three layer model of the human skin. The same analyses was performed in our previous work with conventional RF communication and proposed TR (Thermal Radiation) mode with continuous transmission and reception of data [17]. The temperature elevation recorded in that case was around 1.6-1.9 degree Celsius for current scenario (here Active Mode) and 1.2-1.4 degree Celsius for proposed TR mode. These are denoted by "A" and "B" in Fig.6. The same analysis is performed for our proposed Hybrid mode and pure VLC communication. In the case of Hybrid mode it can be seen from the graph that the temperature elevation produced is around 0.6 to 0.8 degree Celsius. In the case of pure VLC communication there is least temperature elevation 0.2 to 0.4 degree Celsius recorded.

This is denoted by "C" and "D" in Fig.6 and is a result of high energy efficiency in hybrid mode and pure VLC communication. This results in less incident power value in the exposed tissue and hence less temperature elevation produced. We also calculate the performance of the system at the device level i.e. the User Equipment (UE). The performance is calculated in terms of battery life (in hours) of a mobile device communicating in each of the schemes. The device parameters are the same as explained in previous section with the relevant equations to compute the battery lifetime. The battery life of a device operating in pure VLC and pure Wi-Fi is computed w.r.t distance. As the distance increases the battery life of the device decreases.

This is because there is more consumption of dynamic power by the device to maintain a stable connection along with achieving the desired Quality of Service. As the efficiency of VLC communication is higher than that of Wi-Fi, the battery life of the device is also higher. It can be seen from the graph (Fig. 7) that the battery life is almost same at 120m for both VLC and Wi-Fi communication. The second graph in Fig.7 depicts the battery life enhancement due to handover in VLC. As we proposed earlier that handover must take place at 80m distance in case of VLC communication, the battery life also improves after handover significantly. This is because the reduction in distance between the transmitter and receiver requires less consumption of battery power.

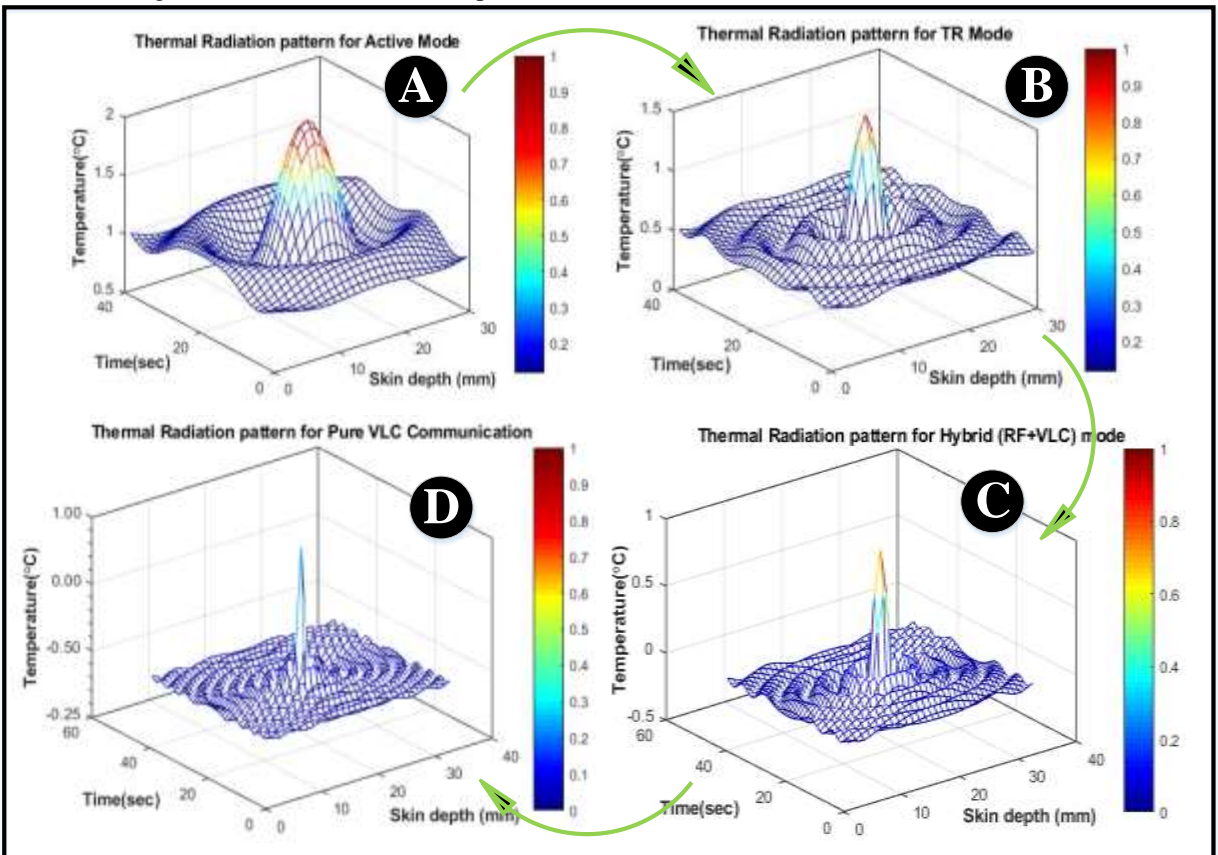

Fig. 6 Comparative analysis of 3D radiation pattern for temperature rise.

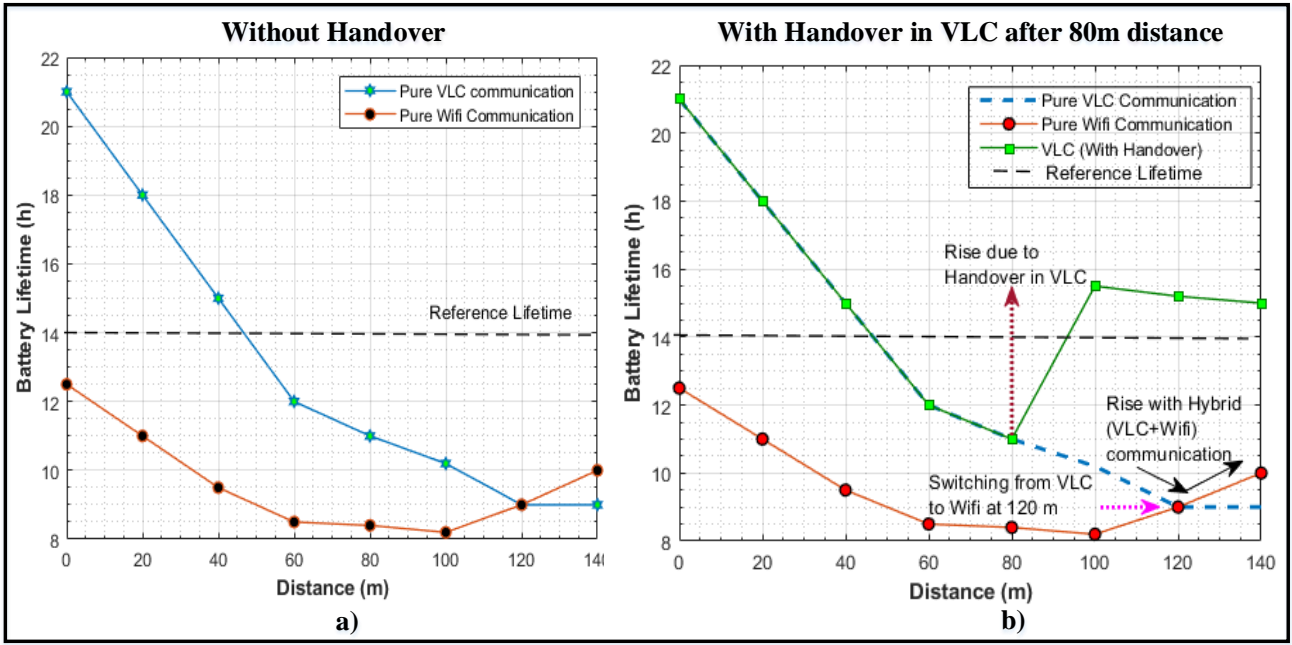

Fig. 7 Battery life v/s Distance for VLC and WiFi Communication a) Without Handover b) With Handover.

For hybrid communication, there is switching from VLC to Wi-Fi at 120m which improves the battery life of a device as there is battery life enhancement when there is a vertical handover performed from VLC to Wi-Fi after 120m.

Another EM radiation metric that is studied to find out the EM radiation impact is SAR i.e. Specific Absorption Rate. The SAR is studied w.r.t skin depth for VLC and Wi-Fi at different distances. The simulations are performed without handover and with handover in VLC. The graphs obtained are shown in Fig.8. The first graph in the figure depicts VLC and Wi-Fi communication performed at 30m and 80m distance respectively. When we compare the values of SAR obtained at 30m distance it can be seen that less SAR is obtained in case of VLC communication and much higher for Wi-Fi. The same pattern is observed at 80m distance (second graph in Fig. 8) for both VLC and Wi-Fi with SAR decreasing as the skin depth increases. The value of SAR decrease as we go deep in to the skin because the outermost layer i.e. Epidermis has a low water content.

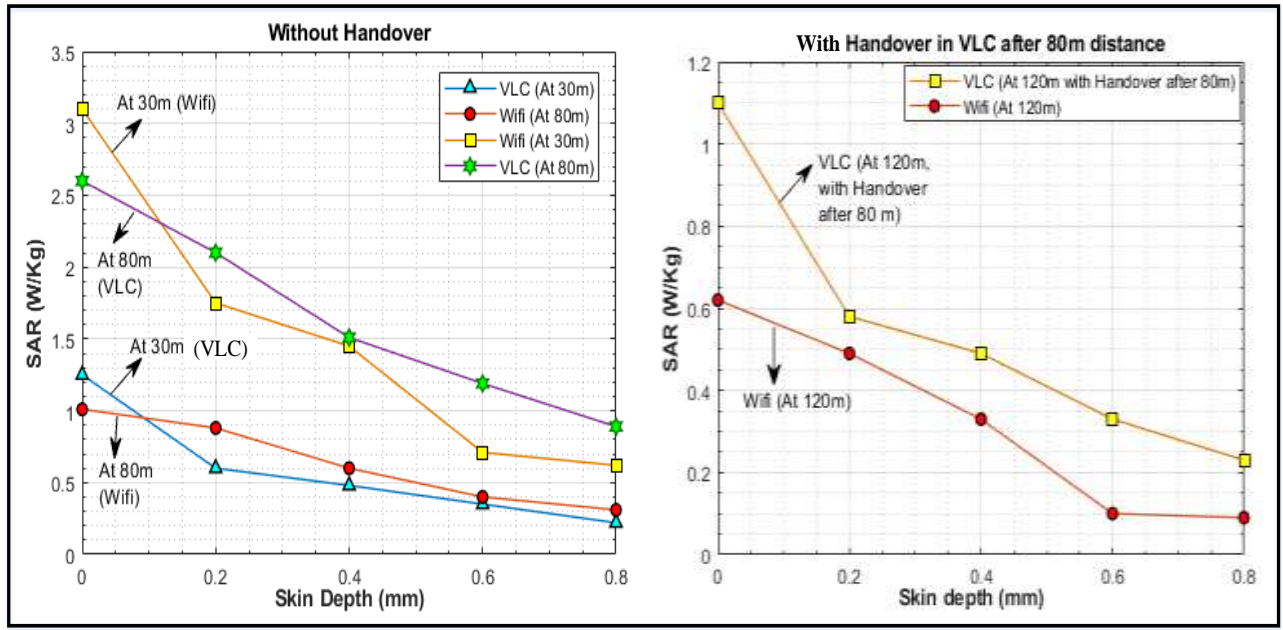

Fig. 8 SAR v/s Skin depth in VLC (With and Without Handover) and WiFi communication.

As we go deeper in the human skin tissue at about 0.06-2.8mm, which is the thickness of epidermis and dermis layers, the water content increase by 70-80% [19]. When the power is incident on the skin, most of the energy is absorbed in the regions with large water concentration hence producing less SAR. Hence it is evident that incorporating hybrid approach for indoor communication is safe and reliable for humans as there is less SAR levels in comparison to RF communication.

A complexity analyses have been performed for Active Mode i.e. Current communication scenario, TR mode, proposed Hybrid mode and Pure VLC (Fig. 9). The comparison has been drawn on the basis of power consumed and power saved by the devices operating in each of the respective modes. It is found that highest percentage of complexity exists when a device is communicating in Active mode. Also as the distance increases the percentage complexity increases because the amount of power required for carrying out the communication in uplink as well as downlink rises. It is evident from the plot that Hybrid (RF + VLC) and pure VLC reduce the overall complexity. At a distance of about 60 m the percent complexity obtained for AM is 62.5% and 41% for pure VLC. Hence there is significant reduction in complexity obtained.

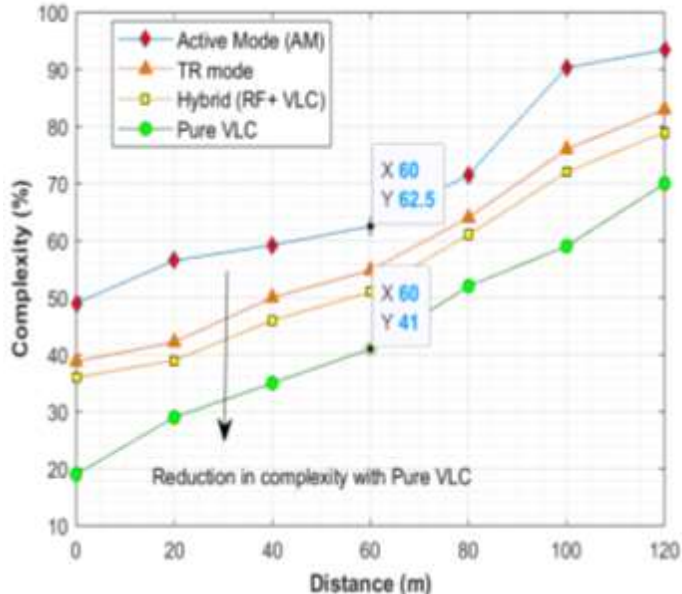

Fig. 9 Comparative analyses of Complexity v/s Distance.

Another EM radiation metric i.e Power density is studied for Wi-Fi and VLC communication w.r.t skin depth. The incident power density is computed w.r.t to the transmitting antenna for a particular mode. The absorbed power density is computed w.r.t Specific absorption rates in the exposed tissue (Fig. 10). Since the power absorption rate is maximum in the epidermis and the dermis layers of the skin tissue, high power density is observed till a skin depth of 0.2 mm and reduces significantly as the depth increases. This is due to the water that water content increases in the deeper layers and hence the absorbed heat is dissipated. High values for incident and absorbed power density are obtained for Wi-Fi and much lower values for VLC. This is because visible light communication does not emit electromagnetic radiation and the rise in power density occurs due to temperature elevation due to heat produced in the exposed tissue over a time period. The lower power density levels with VLC make it very safe for human operation as the eyes, skull and the outermost skin tissue are most susceptible to damage incurred from the harmful EM radiations.

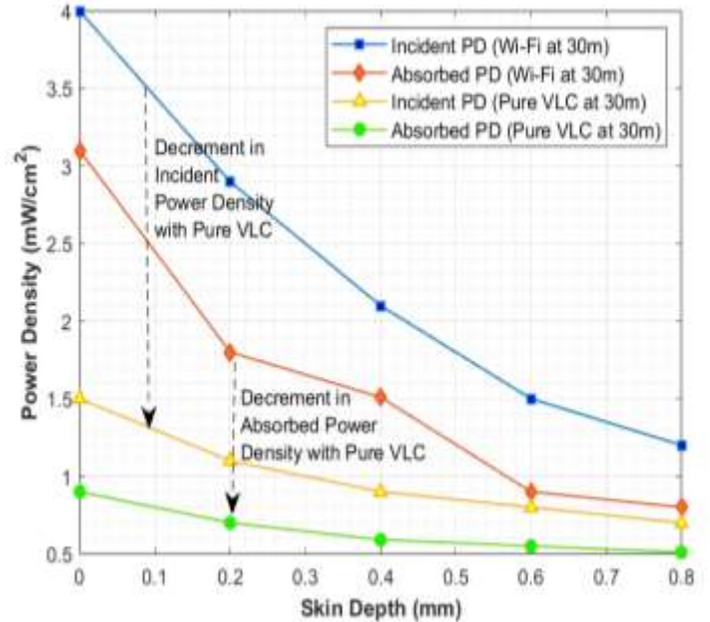

Fig. 10 Comparative Analysis of Incident and Absorbed power density v/s Skin depth.

VI. CONCLUSION

With the advancement in the wireless and mobile communication networks there is a need to make the future networks safe and reliable in operation. This requires a need to revise the existing regulations and guidelines so that the overall EM radiation impact caused due to wireless communications is reduced. The deployment of future communication networks requires certain optimization techniques that can lessen the harmful radiation impact whilst achieving the QoS and QoE required. In this regard, we propose the hybrid communication technique integrating RF and VLC for indoor wireless access. VLC communication can be used indoors as it provides illumination and wireless data transmission simultaneously. The performance of hybrid communication is compared with Wi-Fi communication in terms of parameters such as SINR, Energy efficiency and Battery lifetime. The EM radiation impact for the two is also compared in terms of metrics such as temperature elevation, Specific Absorption Rate, power density and complexity. It is validated from the simulation results obtained that the hybrid communication technique performs better than Wi-Fi communication in an indoor scenario. There is reduction in absorbed power density by 48% after switching from Wi-Fi to VLC. The temperature elevation is almost negligible with VLC as it does not emit harmful EM radiations and SAR reduction obtained is about 59.6%. This makes the hybrid communication mode very safe in operation and more reliable. The proposed technique endorses Green communication as it helps in reducing the carbon footprint in the atmosphere by making the overall network more energy efficient.

ACKNOWLEDGMENT

The authors gratefully acknowledge the support provided by Satish Dhawan Centre for Space Sciences, ISRO, Central University of Jammu, Jammu and Kashmir, India. and Communication Lab at IIIT Jabalpur, India.

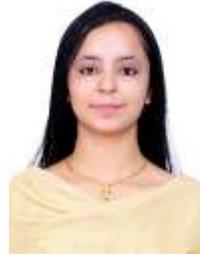
**HANEET KOUR** is currently working as Research Associate at SDCSS, ISRO, Central University of Jammu, J & K. She has completed her Ph. D degree in Electronics and Communication Engineering from Shri Mata Vaishno Devi University, Katra, Jammu and Kashmir, India. Her research interest includes the emerging technologies of 5G/6G wireless communication networks, power optimization, Green Communication, Network Safety/ Reliability and Satellite Communication. She is working on MATLAB tools for Wireless Communication, has received student travel grant from COMSNET in 2019 and 2020. She is a student member of Institute of Electrical and Electronics Engineers (IEEE).

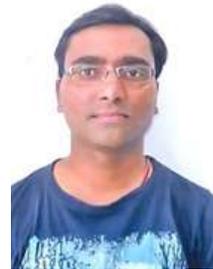
Dr. RAKESH K JHA (S'10, M'13, SM 2015) is currently an Associate Professor in the Department of Electronics and Communication Engineering, Indian Institute of Information Technology Design and Manufacturing Jabalpur, India. He has also worked as an Associate Professor at SMVD University, J&K, India. He is among the top 2% researchers of the world. He has published more than 71 SCI Journals Papers including many IEEE Transactions, IEEE Journal, and more than 25 International Conference papers. His area of interest is Wireless communication, Optical Fiber Communication, Computer Networks, and Security issues. Dr. Jha's one concept related to the router of Wireless Communication was accepted by ITU in 2010. He has received the young scientist author award by ITU in Dec 2010. He has received APAN fellowship in 2011, 2012, 2017 and 2018 and student travel grant from COMSNET 2012. He is a senior member of IEEE, GISFI and SIAM, International Association of Engineers (IAENG) and ACCS (Advance Computing and Communication Society). He is also a member of, ACM and CSI, with many patents and more than 6501 citations to his credit.

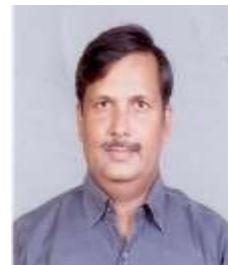
PROF. SANJEEV JAIN**,** born at Vidisha in Madhya Pradesh in 1967, obtained his Post Graduate Degree in Computer Science and Engineering from the Indian Institute of Technology, Delhi, in 1992. He later received his Doctorate Degree in Computer Science & Engineering and has over 24 years' experience in teaching and research. He has served as Director, Madhav Institute of Technology and Science (MITS), Gwalior. and as Director IIITDM, Jabalpur. He has also served at SMVDU, Katra as vice-chancellor. Presently he is serving as Vice- Chancellor at Central University of Jammu, J & K.  Besides teaching at Post Graduate, level Professor Jain has the credit of making significant contributions to R & D in the area of Image Processing and Mobile Adhoc Network.